\documentclass[prd,aps,eqsecnum,floatfix,nofootinbib,preprint,tightenlines]{revtex4}

\usepackage{latexsym}
\usepackage{graphicx}
\usepackage{multirow}
\usepackage{color}
\usepackage{amsmath}
\usepackage{hyperref}  

\def\floatcaption#1#2{ \caption{#2 \label{#1}} }

\def\NON{\nonumber\\}
\def\bibi{\bibitem}

\def\a{\alpha}
\def\b{\beta}
\def\c{\chi}
\def\d{\delta}
\def\e{\epsilon}                
\def\g{\gamma}

\def\j{\psi}

\def\l{\lambda}
\def\m{\mu}
\def\n{\nu}

\def\p{\pi}                     
\def\r{\rho}                    
\def\s{\sigma}                  
\def\t{\tau}

\def\z{\zeta}

\def\G{\Gamma}
\def\J{\Psi}
\def\L{\Lambda}

\def\S{\Sigma}
\def\U{\Upsilon}

\def\cc{{\cal C}}
\def\cd{{\cal D}}

\def\cg{{\cal G}}

\def\cl{{\cal L}}

\def\cp{{\cal P}}

\def\cbo{{\,\raise-.15ex\Sc [\,}}                       

\def\svev#1{\left\langle #1\right\rangle}       

\def\ddt#1{{\buildrel {\hbox{\LARGE .\kern-2pt.}} \over {#1}}}

\def\eg{\mbox{\it e.g.}}

\def\tr{{\rm tr}\,}

\def\half{{1\over 2}}

\def\ttl#1{{\it #1}}

\long \def \blockcomment #1\endcomment{}

\def\tr{\,{\rm tr}}

\def\tJ{\tilde\J}

\def\teta{\tilde\eta}

\def\bB{\overline{B}}
\def\bq{\overline{q}}

\def\bz{\overline\zeta}

\def\bU{\bar{\U}}
\def\bc{\overline{\c}}
\def\bj{\overline{\j}}
\def\bJ{\bar{\J}}
\def\btJ{\bar{\tilde\J}}

\def\bteta{\overline{\tilde\eta}}
\def\etabar{\overline\eta}

\def\one{\mbox{\bf 1}}
\def\three{\mbox{\bf 3}}
\def\threebar{\mbox{${\bf \overline{3}}$}}
\def\five{\mbox{\bf 5}}
\def\fivebar{\mbox{${\bf \overline{5}}$}}
\def\chir{\mbox{\bf s}}
\def\chirbar{\mbox{${\bf \overline{s}}$}}

\def\hs#1{\hspace{#1 ex}}

\def\irrep{\textit{irrep}}
\def\irreps{\textit{irreps}}
\def\idn{\hspace{.5ex}I\hspace{.5ex}}
\def\SU{\mbox{SU}}
\def\SO{\mbox{SO}}
\def\Sp{\mbox{Sp}}
\def\Uone{\mbox{U(1)}}

\begin{document}

\begin{center}
{\large\bf One-loop anomalous dimension of top-partner\\[2mm]
  hyperbaryons in a family of composite Higgs models}\\[8mm]
Thomas DeGrand$^a$ and  Yigal Shamir$^b$\\[8mm]
{\small
$^a$ Department of Physics,
University of Colorado, Boulder, CO 80309, USA\\
$^b$ Raymond and Beverly Sackler School of Physics and Astronomy,\\
Tel~Aviv University, 69978, Tel~Aviv, Israel}\\[10mm]
\end{center}

\begin{quotation}
An important ingredient of many composite-Higgs scenarios is
partial compositeness: a mass for the top quark is generated via mixing with
baryonic operators of the same new strong dynamics that produces
the composite Higgs particle. These baryonic operators are scale dependent.
We construct these operators, and calculate their
one-loop anomalous dimension, in a set of ultraviolet completions
of the Standard Model proposed by Ferretti and Karateev.
\end{quotation}

\newpage
\section{\label{intro} Introduction}
A common theme in the phenomenology of the electroweak interaction is
that the Standard Model (SM) emerges as a low energy effective theory of
some high scale dynamics. That dynamics could be strongly interacting,
with the Higgs boson appearing as a composite Nambu-Goldstone boson (NGB).
The strongly coupled sector is presumed
to have a global symmetry group $G$ which spontaneously breaks to a
subgroup $H$ that, in turn, can accommodate the electroweak symmetries
$\SU(2)_L\times\Uone_Y$.  Turning on electroweak interactions
with the NGBs and, in addition, allowing them
to interact with the top quark converts some of the NGBs into pseudo
Nambu-Goldstone bosons (pNGBs).  The NGB manifold
$G/H$ contains a field with the quantum numbers of the SM Higgs,
which then develops an expectation value, thereby inducing
electroweak symmetry breaking.  These models are generically known as
``composite Higgs models.''
For reviews that reflect the evolution of this field see
\cite{Perelstein:2005ka,Contino:2010rs,Bellazzini:2014yua,Panico:2015jxa}.

We will call the new strong dynamics hypercolor, to distinguish it
from the classic technicolor scenario, and denote by $\L_{HC}$ its scale.
$\L_{HC}$ must be larger than the electroweak scale of 245 GeV,
and phenomenological expectations place it in the range of 1 to a few TeV.
The SM Higgs also gives masses to fermions.
In all composite Higgs extensions of the SM we are familiar with,
this occurs through higher dimensional operators, so that the new interactions
are (formally) irrelevant.  The most important of them are dimension-six,
four-fermion operators.  They are associated with a yet higher
energy scale $\L_{EHC}$ [for extended hypercolor (EHC)], at which, presumably,
another gauge symmetry undergoes spontaneous symmetry breaking,
much like the exchange of $W^\pm$ and $Z$ bosons manifests itself
at the hadronic scale through effective four-fermion operators.

According to one possible mechanism for generating fermion masses,
called ``partial compositeness,'' an elementary SM fermion
couples linearly to strong-sector baryons, which, in effect,
allows it to couple to the composite Higgs as well \cite{Kaplan:1991dc}.
Mass eigenstates are then linear superpositions of SM fermions
and hyperbaryons.
Since it is the only fermion whose mass is comparable to the electroweak scale,
the top quark plays a special role.  Indeed, in the composite-Higgs models
of Ref.~\cite{FerKar} that we will study in this paper, the top's contribution
to the Higgs effective potential is crucial in order to generate a negative
curvature at the origin, and, thus, a non-zero expectation value
for the Higgs field \cite{ferretti,topsect}.

Schematically, each four-fermion interaction is given by
\begin{equation}
  \cl(\m) =  \frac{\l(\m) \bq B + \l^*(\m) \bB q}{\L_{EHC}^2} \ ,
\label{EHC}
\end{equation}
where $q,\bq$ are SM fermions, while $B,\bB$ are interpolating fields
for some three-fermion bound states of the hypercolor theory.
The coupling $\l(\m)$ is dimensionless, and $\m$ is the scale at which
the interaction is probed.  We presume that $\cl(\m)$ was induced
by the exchanged of superheavy gauge bosons
with mass $\sim \L_{EHC} > \L_{HC}$, so that, when probed at that scale,
$\l(\L_{EHC})$ would be equal to the corresponding gauge coupling squared,
up to some $O(1)$ number.

The basic tension confronting such scenarios is that,
generically, the EHC interaction will also induce four-fermion operators
$\sim \bq q \bq q$, that involve only SM fermions.
Flavor physics constraints on such interactions can be quite severe.
Depending on exactly what is assumed about the EHC theory,
meeting experimental bounds may require the scale $\L_{EHC}$ to be
much larger than $\L_{HC}$.  Then the question becomes, can
the observed fermion masses, and, in particular, the top-quark mass,
be generated by interactions such as~(\ref{EHC})?
In order to determine $\l(\L_{HC})$,
the strength of the interaction~(\ref{EHC}) at the hypercolor scale,
we must track the renormalization-group evolution
of the hyperbaryon operator $B$ from $\L_{EHC}$ down to $\L_{HC}$.
This evolution is driven by the same
hypercolor interaction that holds together the constituents of the hyperbaryon.
We will assume that all the SM
gauge interactions, including QCD, are sufficiently weak at the relevant scales
so that their contribution to the renormalization-group flow can be neglected.

So finally, we come to the point of this short note:
first, to write down a set of candidate dimension-six operators
which couple the top quark to hyperbaryons, and then to compute
their anomalous dimensions in the hypercolor theory to one loop.

Which hyperbaryon operators can couple linearly to the top quark,
and how they evolve, are questions that can only be addressed
within a concrete hypercolor model.
In this paper we will study a family of models
proposed by Ferretti and Karateev \cite{FerKar}.
The requirements underlying their classification include:
\begin{itemize}
\item The hypercolor gauge group is simple and asymptotically free.
\item There are no gauge anomalies.
\item The unbroken global symmetry $H$ is such that
\begin{eqnarray}
\label{dynsym}
  H &\supset& \SU(3)_{\rm color}\times \SU(2)_L\times \SU(2)_R\times \Uone_X
\\
    &\supset& \SU(3)_{\rm color}\times \SU(2)_L\times \Uone_Y\ ,
\nonumber
\end{eqnarray}
with the gauge group of the SM in the last line.
The group $\SU(2)_R$ is the familiar custodial symmetry of the SM.
Hypercharge is $Y=T^3_R+X$.
\item The coset $G/H$ contains a field with the quantum numbers
of the SM Higgs.
\item There exist hypercolor baryons with quantum numbers that allow
them to couple linearly to SM fermions.
\end{itemize}
In these models, the Higgs field is identified with pNBGs that
originate from the condensation of fermions in a real or pseudoreal irreducible representation (\irrep)
of the hypercolor gauge group.  The classification of symmetry breaking
patterns was given long ago in Ref.~\cite{Peskin,Preskill:1980mz}.
In the case of a real \irrep, the minimal number of Weyl (or Majorana)
fermions needed to satisfy all requirements is 5.
The symmetry-breaking manifold is then $\SU(5)/\SO(5)$,
with the Higgs field taking up 4 out of a total of 14 pNGBs.
For a pseudoreal \irrep, 4 Weyl fermions are needed.  The coset
is $\SU(4)/\Sp(4)$, with just 5 pNGBs altogether, where again 4 of them
constitute the Higgs field.

In order to have hyperbaryons that can couple linearly to the top quark,
fermions in at least one more \irrep\ of the hypercolor gauge group are needed.
A list of models that satisfy all these requirements was given
in Ref.~\cite{FerKar}.  One model (argued to be the preferred one by Ferretti in
Ref.~\cite{ferretti}) has an $\SU(4)$ gauge group, 5 Majorana fermions in the
two-index antisymmetric (as2) representation,
and 3 fundamental representation Dirac fermions.
Other candidates include $\SO(d)$ gauge fields and a mix of vector and spinor
representations.
(Other proposals for ultraviolet completions which we are aware of include
$\Sp(2N)$ gauge groups and two representations of fermions
\cite{Barnard:2013zea}
and $\SU(3)$ with many fundamentals \cite{Vecchi:2015fma}.)
Each model is minimal
in the sense that, if any fermions are removed, then it would not be possible
to satisfy some of the requirements listed above.  Of course,
additional fermions could be added that do not play any direct role
in meeting the above requirements.  The main effect of such additional matter
would be to slow down the running of the hypercolor coupling.

With only the minimal fermion content listed in Ref.~\cite{FerKar},
many of the models appear to be QCD-like: their beta function is not small,
and they are not expected to exhibit near-conformality.

(This statement can be justified by looking at the two lowest-order coefficients
of the beta function. QCD-like behavior is expected when they are both
positive, in the conventions of Eqs.~(\ref{betafn}) and~(\ref{b12}).
For the $\SU(4)$ model, we can go further by invoking
large-$N$ scaling as an appropriate
way to compare models. We do this by scaling
the lowest order coefficients, dividing by $N$ and $N^2$, $b_1/N$ and $b_2/N^2$.
They would compose the beta function for the 't Hooft coupling $g^2N$.
The $\SU(4)$ Ferretti-Karateev model
has $b_1/4=7/3$, $b_2/16=461/192\simeq 2.4$.
In contrast, $\SU(3)$ gauge theory with $N_f=6$ fundamental Dirac fermions
has $b_1/3= 7/3$, $b_2/9= 26/9 \simeq 2.9$,
equal running at one loop and slightly faster running at two loops.)

In the one-loop approximation, the enhancement (or suppression) factor
resulting from running from a high scale $\m$ down to a smaller scale $\m'$
is then
\begin{equation}
  \frac{\l(\m')}{\l(\m)} =
  \left(\frac{g^2(\m')}{g^2(\m)}\right)^{\frac{\g_1}{2b_1}}
  = \left(\frac{\log(\m/\L_{HC})}{\log(\m'/\L_{HC})}\right)^{\frac{\g_1}{2b_1}} \ ,
\label{RG}
\end{equation}
where now $\L_{HC}$ is the scale occurring the solution
of the beta function\footnote{%
  See Eq.~(\ref{b12}) for the general form of the one- and two-loop coefficients.
}
\begin{equation}
  \b(g^2) = \m \frac{\partial g^2}{\partial\m}
  = -\frac{g^4}{16\p^2}\, b_1 -\frac{g^6}{(16\p^2)^2}\, b_2 + \cdots \ ,
\label{betafn}
\end{equation}
within the one-loop approximation.
The anomalous dimension is
\begin{equation}
  \g(g) = \frac{\m}{Z_B} \frac{\partial Z_B}{\partial \m}
  = -\frac{g^2}{16\p^2}\, \g_1 + \cdots \ ,
\label{gamma}
\end{equation}
where $Z_B$ is the wave-function renormalization factor for the
hyperbaryon $B$.
We use similar conventions for the $\b$ and $\g$ functions.
A negative value for $\g$ (or a positive one-loop coefficient $\g_1$),
means that the coupling $\l(\m')$ grows when $\m'$ is decreased.
For the running from $\L_{EHC}$ down to $\L_{HC}$ we take $\m\approx\L_{EHC}$
and $\m'\approx\L_{HC}$, leading to
\begin{equation}
  \frac{\l(\L_{HC})}{\l(\L_{EHC})} \approx
  \log^{\frac{\g_1}{2b_1}}(\L_{EHC}/\L_{HC}) \ .
\label{logenhnc}
\end{equation}

As mentioned above, additional matter could be added to the model
to slow down the running of the hypercolor coupling.  Suppose that,
all the way from $\L_{EHC}$ to $\L_{HC}$, the hypercolor coupling
was approximately constant and equal to $g_*$.
Then, instead of Eq.~(\ref{logenhnc}), running  would give a power law enhancement
\begin{equation}
  \frac{\l(\L_{HC})}{\l(\L_{EHC})} \approx
\left( \frac{ \L_{EHC}}{\L_{HC}}\right)^{-\g(g_*)} \ ,
\label{enhnc1}
\end{equation}
or, if the one loop formula were valid,
\begin{equation}
  \frac{\l(\L_{HC})}{\l(\L_{EHC})} \approx
\left( \frac{\L_{EHC}}{\L_{HC}} \right)^{\g_1 g_*^2/(16\pi^2)} \ ,
\label{enhnc}
\end{equation}
potentially producing a much bigger effect.

In this paper we construct the top-partner hyperbaryon operators
for some of the models of Ref.~\cite{FerKar}, and compute their one-loop
anomalous dimension.  In Sec.~\ref{SU4} we discuss the $\SU(4)$ model,
and in Sec.~\ref{SOd} we extend the discussion to models based on
an $\SO(d)$ gauge group.  We discuss our results in Sec.~\ref{disc}.
Technical details are relegated to appendices.

The technology needed for this calculation was originally developed
for the study of proton decay in Grand Unified theories \cite{BEGN}.
A comprehensive review may be found in Ref.~\cite{Buras}.

\section{\label{SU4} Results for the SU(4) model}
We review the construction of top-partner hyperbaryon operators,
following Refs.~\cite{ferretti,topsect}.
The hypercolor gauge group is \SU(4).  The fermion content is the following:
We have 5 Majorana fermions $\c_i$, $i=1,\dots,5$, in the real,
two-index antisymmetric \irrep,
and three fundamental representation Dirac fermions.

The Majorana field $\c$ can be written in terms of a
right-handed Weyl fermion $\U$ as
\begin{subequations}
\label{maj}
\begin{eqnarray}
  \c_{ABi} &=&
  \left(
  \begin{array}{c}
    \vspace{1ex}
    \U_{ABi} \\
    \half\e_{ABCD}\,\e\,\bU_{CDi}^T
  \end{array}
  \right) \ ,
\label{maja}\\
  \bc_{ABi} &=& \half \e_{ABCD}\c^T_{CDi} \, C \ = \ \rule{0ex}{3.5ex}
  \left(
  \begin{array}{cc}
    \! -\half\e_{ABCD}\U_{CDi}^T \e \  &  \bU_{ABi}
  \end{array}
  \right) \ .
\label{majb}
\end{eqnarray}
\end{subequations}
We use capital letters for the $\SU(4)$ hypercolor indices.
Multiple indices of a single object will always be fully antisymmetrized.
We have suppressed spinor indices.  $C$ is the charge-conjugation matrix,
$\e=i\s_2$ is the two-dimensional $\e$-tensor acting on the Weyl
spinor index, and the superscript $T$ denotes the transpose in spinor space.

The three Dirac fermions $\j_a$, $a=1,2,3$, are
in the fundamental representation.  $\j_a$ can be written in terms
of two right-handed Weyl fermions, $\J_a$ in the fundamental \irrep\
and $\tJ_a$ in the anti-fundamental, as
\begin{equation}
  \j_{Aa} = \left( \begin{array}{c}
    \J_{Aa} \\ \e\, \btJ^T_{Aa}
  \end{array} \right) \ , \qquad
  \bj_{Aa} = \left( -\tJ_{Aa}^T \e \ \ \ \bJ_{Aa} \right) \ .
\label{DW}
\end{equation}

The global symmetry is
\begin{equation}
\label{flavorG}
  G = \SU(5)\times \SU(3)\times \SU(3)'\times \Uone_X\times \Uone'\ ,
\end{equation}
with quantum numbers $({\bf 5},{\bf 1},{\bf 1})_{(0,-1)}$ for $\U$;
$({\bf 1},{\bf \bar{3}},{\bf 1})_{(1/3,5/3)}$ for $\J$;
and $({\bf 1},{\bf 1},{\bf 3})_{(-1/3,5/3)}$ for $\tJ$.
The symmetry-breaking pattern is $G\to H$ with
\begin{eqnarray}
\label{flavorH}
  H &=& \SU(3)_{\rm color}\times \SO(5)\times \Uone_X \ .
\nonumber
\end{eqnarray}
The unbroken group $H$ satisfies Eq.~(\ref{dynsym}),
with $\SU(2)_L\times \SU(2)_R\subset\SO(5)$.
The symmetry breaking $\SU(5)\to \SO(5)$ is induced by the Majorana-fermion
condensate $\langle\bc_i\c_j\rangle\propto\d_{ij}$.
Like in QCD, the breaking $\SU(3)\times \SU(3)' \to \SU(3)_{\rm color}$,
where $\SU(3)_{\rm color}$ is the diagonal subgroup, is induced by the condensate
$\langle\bj_a\j_b\rangle\propto\d_{ab}$.  Both condensates also break $\Uone'$.

\begin{table}[t]
\vspace*{3ex}
\begin{center}
\begin{tabular}{ c | clccc | c } \hline
  & $SU(5)$ & $SU(3)\times SU(3)'$ & $SU(3)_c$ & $U(1)_X$ & $U(1)'$ &
\\ \hline\hline
  $\U (\J\J)$      & \five    & $(\threebar,\one)\times(\threebar,\one)\to(\three,\one)$
  & \three    &  2/3 &   7/3 & $B_R$ \\
  $\U (\btJ\btJ)$  & \five    & $(\one,\threebar)\times(\one,\threebar)\to(\one,\three)$
  & \three    &  2/3 & -13/3 & $B'_R$ \\
  $\bU (\bJ\bJ)$   & \fivebar & $(\three,\one)\times(\three,\one)\to(\threebar,\one)$
  & \threebar & -2/3 &  -7/3 & $\bB_R$ \\
  $\bU (\tJ\tJ)$   & \fivebar & $(\one,\three)\times(\one,\three)\to(\one,\threebar)$
  & \threebar & -2/3 &  13/3 &  $\bB'_R$ \\
\hline\hline
  $\bU (\J\J)$     & \fivebar & $(\threebar,\one)\times(\threebar,\one)\to(\three,\one)$
  & \three    &  2/3 &  13/3 & $B_L$ \\
  $\bU (\btJ\btJ)$ & \fivebar & $(\one,\threebar)\times(\one,\threebar)\to(\one,\three)$
  & \three    &  2/3 &  -7/3 & $B'_L$ \\
  $\U (\bJ\bJ)$    & \five    & $(\three,\one)\times(\three,\one)\to(\threebar,\one)$
  & \threebar & -2/3 & -13/3 & $\bB_L$ \\
  $\U (\tJ\tJ)$    & \five    & $(\one,\three)\times(\one,\three)\to(\one,\threebar)$
  & \threebar & -2/3 &   7/3 & $\bB'_L$ \\
\hline\hline
\end{tabular}
\end{center}
\vspace*{-3ex}
\begin{quotation}
\floatcaption{tabHC}{%
Local hyperbaryon operators of the $\SU(4)$ model.
The leftmost column gives the Weyl-fermion
content, and the rightmost column the notation used for the operator.
The remaining columns list the quantum numbers.
}
\end{quotation}
\vspace*{-4.5ex}
\end{table}

Let us now discuss the hyperbaryon operators.  We first require that the only
source of explicit breaking of $\SU(3)\times \SU(3)'$ to $\SU(3)_{\rm color}$
will be the QCD interactions.  This implies that the hyperbaryon
must contain two same-type fundamental fermions, e.g., $\J\J$ or
$\btJ\btJ$, but not $\J\btJ$.  The quantum numbers of the allowed
hyperbaryon operators are listed in Table~\ref{tabHC}.
The explicit form of the unprimed operators in Table~\ref{tabHC} is
\begin{subequations}
\label{bar4}
\begin{eqnarray}
  B_{Ria} &=& -\half \e_{ABCD} \e_{abc}\, P_R\, \c_{ABi}\,
                  \left( \j_{Cb}^T \,C P_R\, \j_{Dc} \right)
\label{bar4a}\\
  &=& \half \e_{ABCD} \e_{abc}\, \U_{ABi}
      \left(\J^T_{Cb}\,\e\,\J_{Dc}\right) \ ,
\nonumber\\
  \bB_{Ria} &=& \half \e_{ABCD} \e_{abc}\, \bc_{ABi} P_L\,
                  \left( \bj_{Cb} \,C P_L\, \bj_{Dc}^T \right)
\label{bar4b}\\
  &=& \half \e_{ABCD} \e_{abc}\, \bU_{ABi}
      \left(\bJ_{Cb}\,\e\,\bJ_{Dc}^T\right) \ ,
\nonumber\\
  B_{Lia} &=& -\half \e_{ABCD} \e_{abc}\, P_L \c_{ABi}\,
                  \left( \j_{Cb}^T \,C P_R\, \j_{Dc} \right)
\label{bar4c}\\
  &=& \e_{abc}\,\e\,\bU_{ABi}^T
      \left(\J^T_{Ab}\,\e\,\J_{Bc}\right) \ ,
\nonumber\\
  \bB_{Lia} &=& \half \e_{ABCD} \e_{abc}\, \bc_{ABi} P_R\,
                  \left( \bj_{Cb} \,C P_L\, \bj_{Dc}^T \right)
\label{bar4d}\\
  &=& \e_{abc}\,\U^T_{ABi}\,\e
      \left(\bJ_{Ab}\,\e\,\bJ_{Bc}^T\right) \ .
\nonumber
\end{eqnarray}
\end{subequations}
The chiral projectors are $P_R = (1+\gamma_5)/2$ and $P_L=(1-\gamma_5)/2$.
The primed operators are obtained from Eq.~(\ref{bar4})
by interchanging $P_R\leftrightarrow P_L$ inside the $\j\j$
and $\bj\bj$ bilinears.  The $CP$ transformation acts as
\begin{equation}
  \j \to \g_2 \bj^T \ , \qquad \bj \to \j^T \g_2 \ ,
\label{CP}
\end{equation}
for both Dirac and Majorana fermions.
The sign choices of Eqs.~(\ref{bar4a})-~(\ref{bar4d})
imply that the hyperbaryon operators will transform in the same way,
and eventually, that the 4-Fermi lagrangian that couples the SM fermions
to the hypercolor baryons will be $CP$-invariant.

\begin{figure}[t]
\begin{center}
\includegraphics*[width=4cm]{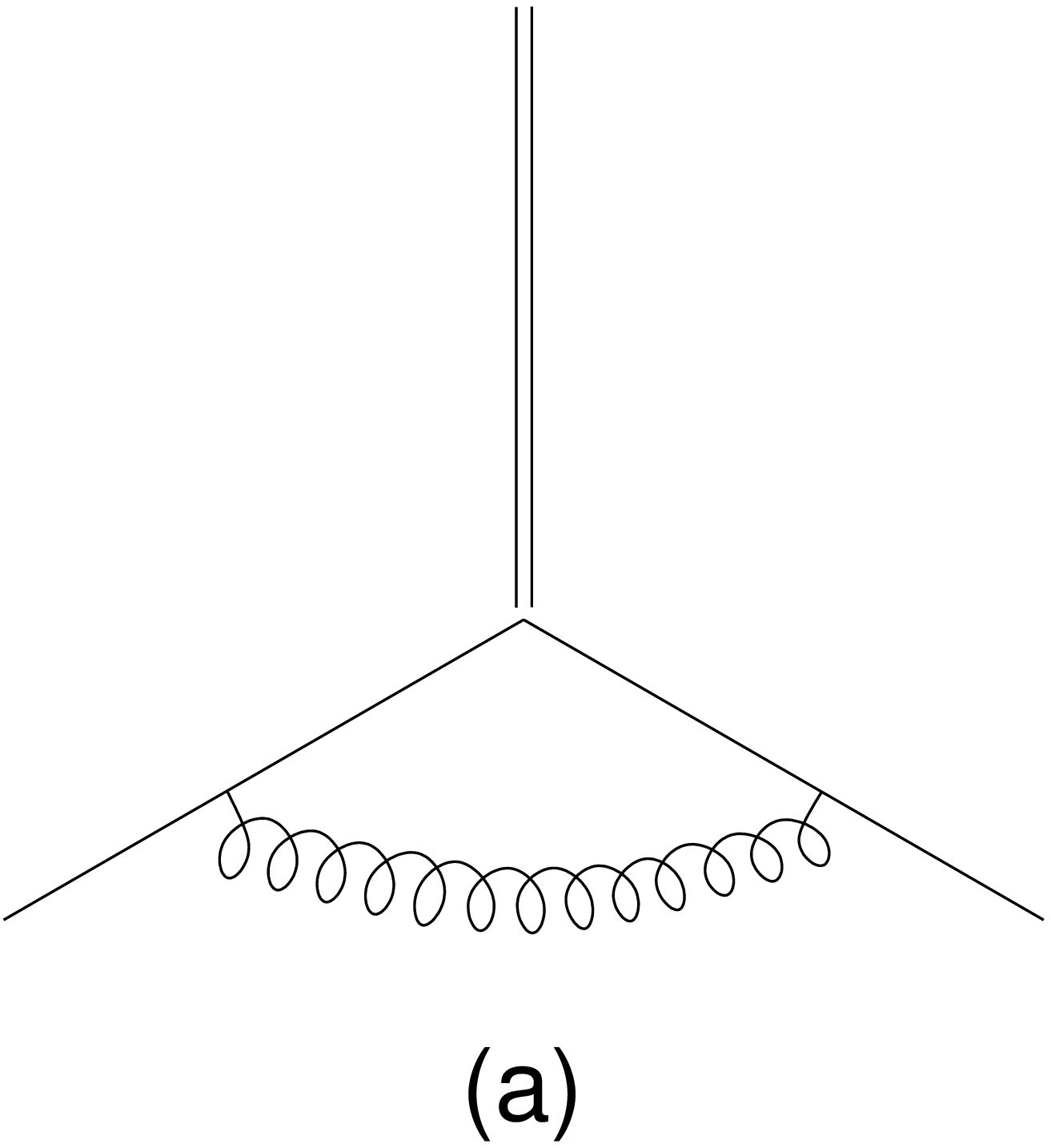}
\hspace{1cm}
\includegraphics*[width=4cm]{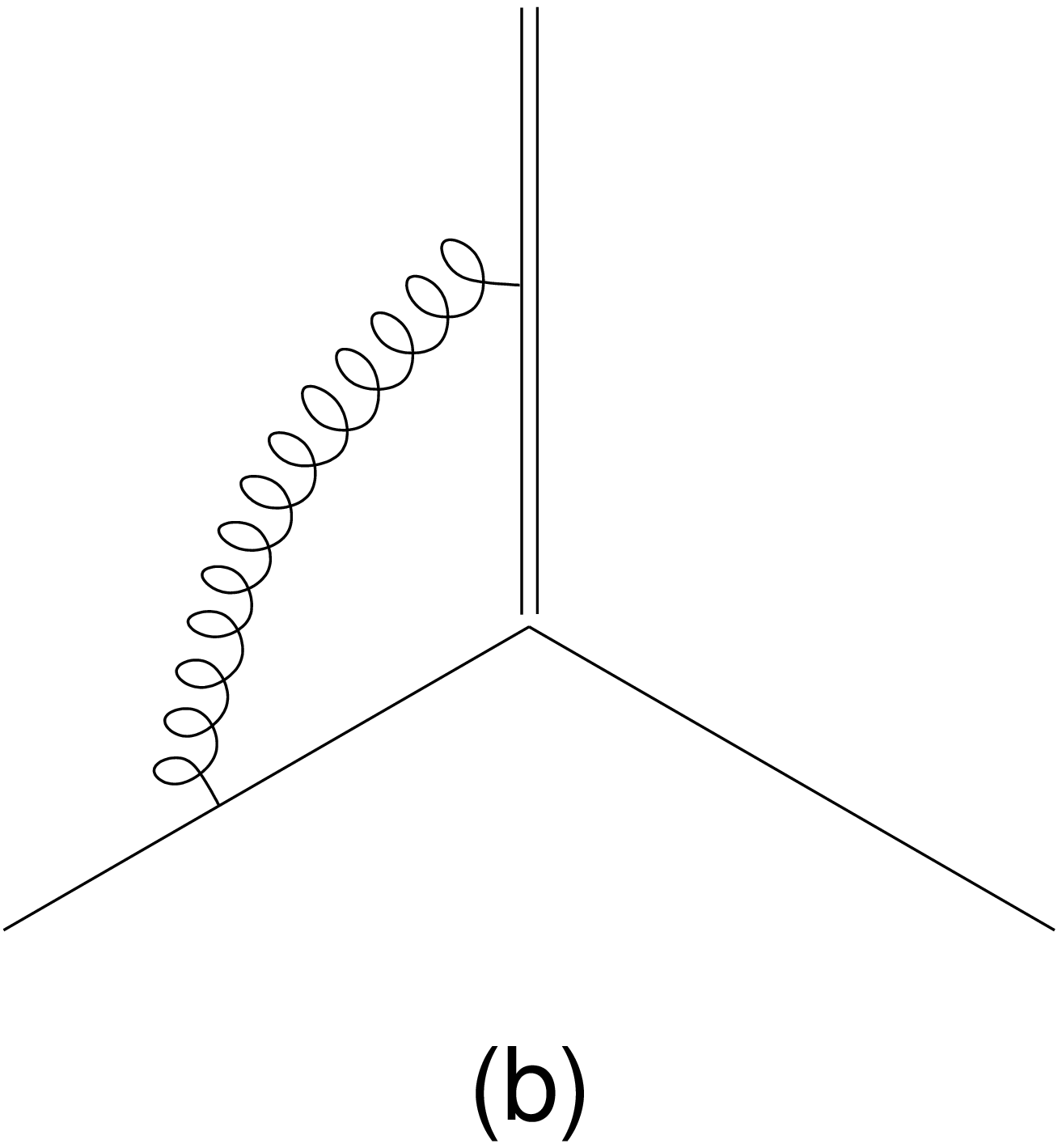}
\hspace{1cm}
\includegraphics*[width=4cm]{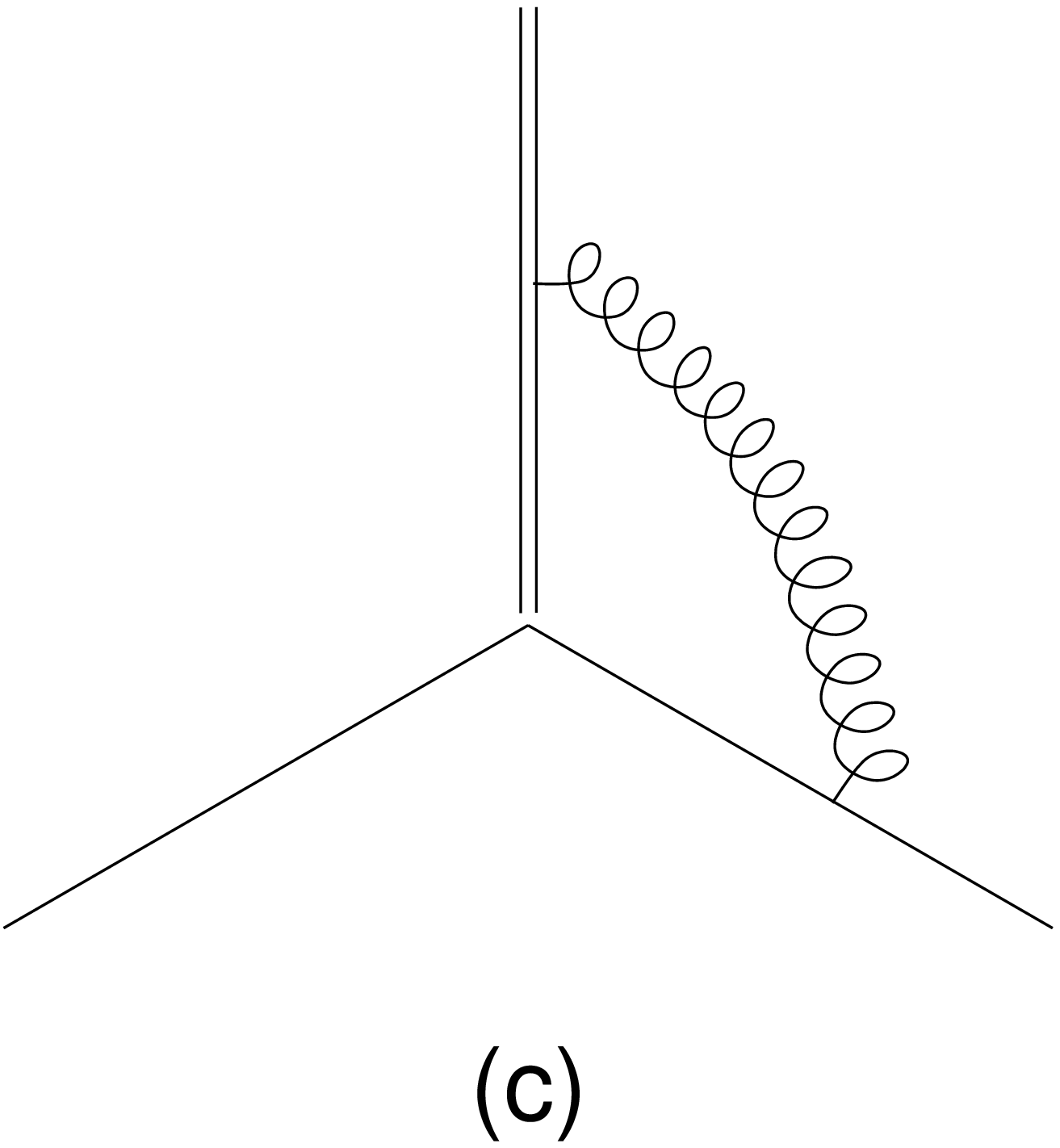}
\end{center}
\begin{quotation}
\floatcaption{diags}%
{One-loop diagrams.
For the $\SU(4)$ model, the double line represents a sextet (as2) fermion,
which carries the open spin index of the hyperbaryon.
The other two lines are fundamental-\irrep\ fermions.
For $\SO(d)$ models, the double line belongs to the vector \irrep,
while the other fermion lines belong to a (chiral) spinor \irrep.
In graph (a) the gluon is exchanged between the two
same-\irrep\ fermions, and in graphs (b) and (c) between
different-\irrep\ fermions.}
\end{quotation}
\vspace*{-4ex}
\end{figure}

At the one-loop level we encounter the three one-particle irreducible diagrams
shown in Fig.~\ref{diags}.  The double line represents the sextet Majorana
fermion, while the single lines represent the two fundamental Dirac fermions.
In all cases studied in this paper the matrix of wave-function
renormalization factors is diagonal.  The $Z$-factor of a given
hyperbaryon operator can be expressed as
\begin{equation}
  Z = 1 + \frac{g^2}{64\p^2\e}
  \Big( \cg(a) \cd(a) + \cg(b,c) \cd(b+c) - \sum_{j=1}^3 C_2(j) \Big)\ ,
\label{Z}
\end{equation}
where $\e=2-d/2$.  With Eq.~(\ref{gamma})
and $\b(g^2,\e)=-2\e g^2 + \b(g^2)$, the one-loop anomalous dimension $\g_1$
is half the expression inside the parentheses on the right-hand side.

In Eq.~(\ref{Z}), the summation in last term
accounts for the wave-function renormalization
factors of the external legs, here given by
$2C_2({\rm fund})+C_2({\rm sextet})$.
After separating out a common factor arising from the logarithmically divergent
momentum integral, the contribution of each one-particle irreducible graph
can be expressed as the product of a group theoretical factor $\cg$
and a Dirac-algebra factor $\cd$.  Fig.~\ref{diags}(a) gives rise
to the term $\cg(a) \cd(a)$.  Figs.~\ref{diags}(b) and~(c) share a common
group theoretical factor that we denote as $\cg(b,c)$,
and their combined contribution is given by $\cg(b,c) \cd(b+c)$,
where $\cd(b+c)$ sums up the Dirac factors from both graphs.
The necessary $\SU(N)$ group theory is summarized in App.~\ref{SUN}.
Details on the calculation of the Dirac factors are listed in  App.~\ref{dirac4}.
The resulting factors are summarized in Tables~\ref{tabdirac} and~\ref{tabclr}.

The operators in Table~\ref{tabHC} all share
the Dirac tensor structure $P_1\otimes P_2$,
where $P_1$ and $P_2$ can each be $P_L$ or $P_R$.
The first projector $P_1$ is sandwiched between the two fundamental fermions:
either between $\j^T C$ and $\j$, or between $\bj$ and $C\bj^T$,
while $P_2$ multiplies the open fermion index of the Majorana fermion $\c$.
The Dirac factors $\cd(a)$ and $\cd(b+c)$ do not depend on the choice
of the chiral projectors.  Thus,
for all the operators in Table~\ref{tabHC} we obtain
\begin{equation}
  \g_1 (P_1 \otimes P_2) = \frac{15}{4} \ ,
\label{g1PP}
\end{equation}
where we have indicated the common Dirac tensor structure of the hyperbaryons.

\begin{table}[thb]
\vspace*{2ex}
\begin{center}
\begin{tabular}{c|rrr} \hline \hline
 & $P_1 \otimes P_2$ & $P_1\g_\m \otimes P_2\g_\m$
 & $\s_{\m\n} \otimes \s_{\m\n}$ \\ \hline
(a)   & 16 \hs{2.5} &  4 \hs{2.5} &  0 \hs{3.5} \\
(b+c) &  8 \hs{2.5} & 20 \hs{2.5} & 24 \hs{3.5} \\ \hline
\hline
\end{tabular}
\end{center}
\begin{quotation}
\floatcaption{tabdirac}{Dirac factors arising from diagram (a)
and from the sum of diagrams (b) and (c), for the various possible
tensor structures that form the hyperbaryon.
$P_1$ and $P_2$ can each be one of the chiral projectors $P_{L,R}$.}
\end{quotation}
\vspace*{-4ex}
\end{table}

If we allow $\SU(3)\times \SU(3)'$ to be broken explicitly
to $\SU(3)_{\rm color}$ by the SM--hypercolor lagrangian,
we may also consider hyperbaryons
where the two fundamental fermions are mixed, \eg, $\btJ\J$.
This allows for two more operators
\begin{equation}
  B^{mix}_{L,R} \propto \e_{ABCD} \e_{abc}\, P_{L,R}\g_\m\, \c_{ABi}\,
                  \left( \j_{Cb}^T \,C \g_\m\g_5\, \j_{Dc} \right) \ .
\label{Bmixed}
\end{equation}
The color factors are the same as before, while the Dirac factors
are shown in the $\g_\m \otimes \g_\m$ column of Table~\ref{tabdirac}.
The resulting one-loop anomalous dimension is
\begin{equation}
  \g_1 (P_1 \g_\m \otimes P_2 \g_\m) = \frac{15}{2} \ ,
\label{g1gg}
\end{equation}
which is twice as big as that of Eq.~(\ref{g1PP}).

Since the spin wave function of the fundamental fermions must be
antisymmetric, the symmetry properties under charge conjugation
(Eq.~(\ref{CCrules})) imply that we can use neither $\j^T C \g_\m\, \j$
nor $\j^T C \s_{\m\n}\, \j$.

\begin{boldmath}
\section{\label{SOd} Results for $\SO(d)$ models}
\end{boldmath}
In this section we study models whose gauge symmetry is the
orthogonal group $\SO(d)$.  We will mainly focus on even values of $d$.
These models, which are discussed in Sec.~\ref{SOdeven},
correspond to the $p=3$ case of Ref.~\cite{FerKar}.
We briefly discuss the $p=2$ models \cite{FerKar}
where the gauge group is $\SO(d)$ with $d$ odd in Sec.~\ref{SOdodd},
and ``flipped'' models where the role of the vector and spinor \irreps\
is interchanged in Sec.~\ref{flip}.  The calculation of the one-loop
anomalous dimension for all these models is summarized in Sec.~\ref{g1SOd}.
For convenience, we review in App.~\ref{apSOd} the needed $\SO(d)$ group theory,
focusing mainly on the spinor \irreps. This Appendix is a bit long,
but it turns out that determining the charge
conjugation properties of our states  is not completely straightforward
until we look at the explicit form of the Dirac matrices.

\bigskip
\begin{boldmath}
\subsection{\label{SOdeven} $\SO(d=2n)$}
\end{boldmath}
Here we consider $\SO(d)$ models with $d=6,8,10,12,14$.
The $d=6$, 8 and 10 models are QCD-like in that $b_1$ and $b_2$
are both positive.
For $d\ge 16$ the hypercolor theory loses asymptotic freedom.
The $d=12$ and 14 models have Banks-Zaks fixed points.
Notice, however, that the number of additional triplets of ordinary color
grows with $d$, and that QCD loses asymptotic freedom already for $d\ge10$.

In all models, the fermion content includes 5 Majorana fermions $\c_i$
in the vector \irrep, which, like the sextet fermions of $\SU(4)$,
produce the coset $\SU(5)/\SO(5)$ after chiral symmetry breaking.
Instead of the $\SU(4)$ fundamental \irrep\ fermions we will now have two
color triplets of Weyl fermions each belonging to some chiral spinor \irrep.
The smallest gauge group $\SO(6)$ is isomorphic to $\SU(4)$.
Hence, this case provides a cross check of the calculation.

\begin{table}[t]
\vspace*{2ex}
\begin{center}
\begin{tabular}{c|cc} \hline \hline
 & $\SU(4)$ & $\SO(d)$ \\ \hline
(a)   & $-\frac{5}{8}$ & $-\frac{(d-2)^2-d}{8}$ \\
(b,c) & $-\frac{5}{4}$ & $-\frac{d-1}{2}$ \\ \hline
\hline
\end{tabular}
\end{center}
\begin{quotation}
\floatcaption{tabclr}{Hypercolor factors arising from diagram (a)
and from diagram (b) or (c).
For some more details on the $\SU(4)$ calculation, see App.~\ref{SUN}.
For the $\SO(d)$ case, see App.~\ref{SOdidns}.
}
\end{quotation}
\vspace*{-4ex}
\end{table}

The gauge singlets that form the hyperbaryons are always constructed
from one $\SO(d)$ vector, and from two $\SO(d)$ spinors
(that also carry the ordinary color).
As we will see, the details of how a vector is formed from two spinors vary.
Our $\SO(d)$ notation is explained in App.~\ref{apSOd}, and
the results pertaining to the Dirac algebra and charge conjugation symmetry
in $d=2n$ dimensions are collected in App.~\ref{dDirac}.
We are guided by Table~\ref{tabSigns}, where we have tabulated the sign
factors that occur in some key relations, which, in turn,
will influence the construction.  We will have to distinguish
several cases.  First, for $d=4n+2$ the chiral spinor \irreps\ are complex
conjugate.  We denote the chiral \irrep\ projected by $\cp_+$ and $\cp_-$
(Eq.~(\ref{Ppm})) as $\chir$ and $\chirbar$ respectively.
We will have to further distinguish between $d=8n+2$ and $d=8n+6$.
The last case is $d=4n$ where the chiral spinor \irreps\ are real (for
$d=8n+4$) or pseudoreal (for $d=8n$).  The \irrep\ projected by
$\cp_\pm$ is denoted $\chir_\pm$ in this case.

\bigskip
\begin{boldmath}
\subsubsection{\label{d8np6} $d=8n+6$}
\end{boldmath}
We start with the $d=8n+6$ case, remembering that $\SO(6)$ should reproduce
the $\SU(4)$ model of Sec.~\ref{SU4}.  The same construction that works
for $\SO(6)$ will work also for $\SO(14)$.  The complex-\irrep\
Weyl fermions are $\J_{a\a}$, transforming as $(\chir,\threebar,\one)$
under $\SO(d)\times \SU(3)\times \SU(3)'$, and $\tJ_{a\a}$, transforming
as $(\chirbar,\one,\three)$.
We use $\a,\b,\ldots=1,2,$ for the 4-dimensional Weyl index,
and, as before, $a,b,\ldots=1,2,3,$ for triplets of $\SU(3)$ or $\SU(3)'$.

Performing the $\SO(d)$ contractions requires some care, because
the spinor conventions which are customary  when $\SO(d)$
is the spacetime symmetry are different from those we will prefer
when $\SO(d)$ is an internal (gauge) symmetry.  We start by grouping
the chiral spinors into $\SO(d)$ Dirac spinors,
\begin{equation}
  \z = \left(\begin{array}{c} \J \\ \tJ \end{array} \right) \ ,
  \qquad  \bz = \left( \bJ \ \ \btJ \right) \ .
\label{SOdirac}
\end{equation}
These objects transform as (see Eq.~(\ref{spingen}))
\begin{equation}
  \d\z = \S_{ij} \z \ , \qquad \d\bz = -\bz\, \S_{ij} \ .
\label{dSOdirac}
\end{equation}
Of course, this implies that $\z,\bz$ span a {\it reducible} representation
of $SO(d)$.  In terms of the irreducible chiral representations,
this is equivalent to
\begin{eqnarray}
  \d\J = (\S_{ij})_{++} \J \ , \qquad \d\bJ =  -\bJ (\S_{ij})_{++} \ ,
\label{dSOchir}\\
  \d\tJ = (\S_{ij})_{--} \tJ \ , \qquad \d\btJ =  -\btJ (\S_{ij})_{--} \ ,
\end{eqnarray}
where we are using the block notation of Eq.~(\ref{block}).
These look like conventional gauge transformation rules,
with generators $(\S_{ij})_{++}$ and $(\S_{ij})_{--}$ for the $\chir$
and $\chirbar$ \irreps\ respectively.  We furthermore observe that,
thanks to the $d$-dimensional charge conjugation symmetry~(\ref{CSig}),
$\cc \bz^T$ and $\z$ have the same $\SO(d)$ transformation properties,
and the same is true for $\bz$ and $\z^T\cc$.

With this information at hand,
we may now construct bilinears that transform as an $\SO(d)$ vector,
and which are constructed from two copies of the same chiral \irrep.
They are
\begin{eqnarray}
  \z^T (\cc \G_i)_{++} \z &\Rightarrow&
    \J^T_{b\a} (\cc \G_i)_{++} \J_{c\b} \ ,
\label{prodJ}\\
  \bz (\G_i \cc)_{++} \bz^T &\Rightarrow&
    \bJ_{b\a} (\G_i \cc)_{++} \bJ^T_{c\b} \ ,
\NON
  \z^T (\cc \G_i)_{--} \z &\Rightarrow&
    \tJ^T_{b\a} (\cc \G_i)_{--} \tJ_{c\b} \ ,
\NON
  \bz (\G_i \cc)_{--} \bz^T &\Rightarrow&
    \btJ_{b\a} (\G_i \cc)_{--} \btJ^T_{c\b} \ ,
\nonumber
\end{eqnarray}
where on the right-hand side we have exposed the chiral \irrep\ content,
and reintroduced the 4-dimensional Weyl index, as well as the triplet index
of $\SU(3)$ or $\SU(3)'$.
According to Table~\ref{tabSigns}, these products are antisymmetric
on their $\SO(6)$ spinor indices, consistent with the contractions of
the fundamental-\irrep\ fermions in the $\SU(4)$ language.

Next, we deal with the 4-dimensional Dirac structure.
To this end, we will group the Weyl fermions into ordinary Dirac fermions
that belong to a specific \irrep, $\chir$,
of the $\SO(d)$ gauge group.\footnote{%
  The 4-dimensional Dirac fermions are obviously different from
  the objects introduced in Eq.~(\ref{SOdirac}), which have components transforming
  as both $\chir$ and $\chirbar$ under $\SO(d)$,
  and are Weyl fermions as far as their spacetime transformation is concerned.
}
Once again using the $\SO(d)$
charge-conjugation symmetry, the Dirac fermions are (compare Eq.~(\ref{DW}))
\begin{equation}
  \j = \left( \begin{array}{c}
    \J \\ \e\, \cc_{+-} \btJ^T
  \end{array} \right) \ , \qquad
  \bj = \left( -\tJ^T \e\, \cc_{-+} \ \ \ \bJ \right) \ .
\label{DiracSOd}
\end{equation}
To be explicit, the $\SO(d)$ transformation rules are
$\d\j = (\S_{ij})_{++}\j$, $\d\bj = -\bj(\S_{ij})_{++}$.
Contracting the 4-dimensional fermion index into an as-yet unspecified
Dirac matrix $X$, the bilinears of Eq.~(\ref{prodJ}) become
\begin{eqnarray}
  \e_{abc}\, \j^T_b (\cc \G_i)_{++} C P_R X P_R \j_c \ ,
\label{prodj}\\
  \e_{abc}\, \bj_b (\G_i \cc)_{++} P_L X P_L C \bj^T_c \ ,
\NON
  \e_{abc}\, \bj_b (\G_i \cc)_{++} P_R X P_R C \bj^T_c \ ,
\NON
  \e_{abc}\, \j^T_b (\cc \G_i)_{++} C P_L X P_L \j_c \ ,
\nonumber
\end{eqnarray}
where we have performed the color contraction as well.
The 4-dimensional Dirac matrix $X$ must commute with $\g_5$.
Moreover, the 4-dimensional spinor wave function must be antisymmetric,
and consulting Eq.~(\ref{CCrules}), the only choice is $X=I$.
As expected, after multiplying by $\c_i$ to form an $\SO(d)$ singlet,
this reproduces the hyperbaryons of Table~\ref{tabHC} and Eq.~(\ref{bar4}).

As in Sec.~\ref{SU4}
we may decide to relax the assumption that $\SU(3)$ and $\SU(3)'$
are good symmetries separately, and insist only on $\SU(3)_{\rm color}$.
This allows us to replace $P_L X P_L$ or $P_R X P_R$
by $\g_\m \g_5$ in Eq.~(\ref{prodj}).

\bigskip
\begin{boldmath}
\subsubsection{\label{d8np2} $d=8n+2$}
\end{boldmath}
This case is represented only by $\SO(10)$.
The construction is similar, except that the products in Eq.~(\ref{prodJ})
are now symmetric on the $\SO(10)$ spinor index.
Hence, the 4-dimensional spin wave function must be symmetric as well.
If we want $SU(3)$ and $SU(3)'$ to be good symmetries separately,
the Dirac fermion bilinears we can use are again given by Eq.~(\ref{prodj}),
but now we must take $X=\s_{\m\n}$.  As before, relaxing this condition
will allow for $X=\g_\m$ as well.  [This time, both
$C P_R \g_\m$ and $C P_L \g_\m$ will be projected to
$C \g_\m$, which is the symmetric spin wave function.]

\bigskip
\begin{boldmath}
\subsubsection{\label{d4n} $d=4n$}
\end{boldmath}
As mentioned already, for $\SO(4n)$ the chiral \irreps\ are real
or pseudoreal.  This means that there would be no hypercolor gauge anomalies
even if we use Weyl fermions that belong only to one of the chiral \irreps.
But, because now $\cc$ commutes with $\cp_\mp$, we have
$(\cc\G_i)_{++}=(\cc\G_i)_{--}=0$, as well as, as always,
$(\G_i)_{++}=(\G_i)_{--}=0$.  A bilinear transforming as an $\SO(4n)$ vector
must therefore be constructed from one \irrep\ of each chirality.
It also follows that we cannot construct hyperbaryons that will respect
$SU(3)$ and $SU(3)'$ separately.  Insisting on $\SU(3)_{\rm color}$ only,
we take $\J_{a\a}$ to be in $(\chir_+,\threebar)$
of $\SO(d)\times \SU(3)_{\rm color}$,
while $\tJ_{a\a}$ is in $(\chir_-,\three)$.
The products that transform as an $\SO(d)$ vector
and $\SU(3)_{\rm color}$ triplets are
\begin{eqnarray}
  && \e_{abc} \btJ_{b\a} (\G_i)_{-+} \J_{c\b} \ ,
\label{prodJ4n}\\
  && \e_{abc} \bJ_{b\a} (\G_i)_{+-} \tJ_{c\b} \ .
\nonumber
\end{eqnarray}
Just in order to be able to use the 4-dimensional Dirac matrices,
we can trivially embed the Weyl fermions into 4-component fermions
with one vanishing chirality,
\begin{equation}
  \J \to \eta = P_R \eta \ , \quad
  \bJ \to \etabar = \etabar P_L \ , \quad
  \tJ \to \teta = P_R \teta \ , \quad
  \btJ \to \bteta = \bteta P_L \ ,
\label{Jeta}
\end{equation}
so that Eq.~(\ref{prodJ4n}) leads to
\begin{eqnarray}
  && \e_{abc}\, \bteta_b (\G_i)_{-+} P_L X P_R \eta_c \ ,
\label{prodj4n}\\
  && \e_{abc}\, \etabar_b (\G_i)_{+-} P_L X P_R \teta_c \ .
\nonumber
\end{eqnarray}
Since we have two different chiral \irreps\ there is
no definite symmetry.  Given the four-dimensional chiral projectors
we take $X=\g_\m$.

This completes the construction of the hyperbaryon operators for $\SO(2n)$.
For convenience, we summarize our findings in Table~\ref{tabops}.
As a final comment, we note that our analysis applies to the minimal
fermion content of these models, and that adding more fermions will in general
lead to more possibilities.  For example, in the $d=4n$ case,
adding new fermions $\J'_{a\a}$ in $(\chir_+,\three)$
and $\tJ'_{a\a}$ in $(\chir_-,\threebar)$ would allow for bilinears
such as $\e_{abc} \tJ_{b\a}^T (\cc\G_i)_{-+} \J'_{c\b}$
and $\e_{abc} \J_{b\a}^T (\cc\G_i)_{+-} \tJ'_{c\b}$.

\begin{table}[t]
\vspace*{2ex}
\begin{center}
\begin{tabular}{c|c|ccc} \hline \hline
 & $d$ & $P_1 \otimes P_2$ & $P_1\g_\m \otimes P_2\g_\m$
 & $\s_{\m\n} \otimes \s_{\m\n}$ \\ \hline
$4n$ & 8,12 & $-$ & B & $-$ \\
$8n+2$ & 10 & $-$ & B & A \\
$8n+6$ & 6,14 & A & B & $-$ \\ \hline
\hline
\end{tabular}
\end{center}
\begin{quotation}
\floatcaption{tabops}{Summary of $\SO(2n)$ hyperbaryon properties.
The first column shows the mod-8 family, and the next the relevant
values of $d$.  For each of the three Dirac tensor structures,
the symbols have the following meaning: A or B indicates that the tensor
structure is allowed for the given family, whereas a $-$ sign indicates
that it is disallowed.  For A, both $\SU(3)$ and $\SU(3)'$ are
good symmetries, whereas for B, only $\SU(3)_{\rm color}$
is a good symmetry. The disallowed cases refer to the minimal fermion
content as described in the main text.}
\end{quotation}
\vspace*{-4ex}
\end{table}

\bigskip
\begin{boldmath}
\subsection{\label{SOdodd} $\SO(d=2n+1)$}
\end{boldmath}
Some of the $p=2$ cases of Ref.~\cite{FerKar} make use of an $\SO(d)$ gauge group
with $d$ odd.  In this case there are no chiral spinor \irreps,
and the Dirac-like \irrep\ is real or pseudoreal.
We take 6 Weyl fermions in the Dirac \irrep, so that their global symmetry
is $\SU(6)$.  We embed
$\SU(3)_{\rm color} \subset \SU(3)\times\SU(3)' \subset \SU(6)$ by declaring
that the first three copies, denoted $\J_{a\a}$, transform as $\three$
of $\SU(3)$, while the last three, denoted $\tJ_{a\a}$,
transform as $\threebar$ of $\SU(3)'$.
As before, we have the 5 Majorana fermions in the vector \irrep.

When constructing the spinor bilinears we have to distinguish between
$d=4n+1$ and $d=4n+3$.  In the latter case, the bilinear
$\e_{abc} \J_{b\a}^T \cc\G_i \J_{c\b}$ transforms as a $d$-dimensional vector,
whereas in the former case this is true for
$\e_{abc} \J_{b\a}^T \G_h\cc\G_i \J_{c\b}$
where $\G_h$ is $\g_5$ or its generalization (see App.~\ref{dDiracodd}).
The hyperbaryon--SM coupling that uses these bilinears will break $\SU(6)$
explicitly to $\SU(3)\times\SU(3)'$.
In addition, we may use the bilinear $\e_{abc} \btJ_{b\a} \G_i \J_{c\b}$
for any odd $d$, in which case only $\SU(3)_{\rm color}$ is preserved.
We leave it to the reader to figure out which 4-dimensional Dirac structures
are allowed by the symmetry properties in each case.

\bigskip
\subsection{\label{flip} Flipped models}
When the spinor \irrep\ is real or pseudoreal, we may interchange
the role of the vector and spinor \irreps.
For example, when the spinor \irrep\ is real, we may use 5 copies
of this \irrep\ to generate the $\SU(5)/\SO(5)$ coset.
In this case, 6 copies of the vector \irrep\ will be used,
and the identification of ordinary color with a subgroup of $\SU(6)$ will be
as in the previous subsection.  When the spinor \irrep\ is pseudoreal,
the minimal number is 4 copies, which trades the $\SU(5)/\SO(5)$ coset
with $\SU(4)/\Sp(4)$.  Once again we leave the detailed
construction of the hyperbaryons to the reader.

\begin{boldmath}
\subsection{\label{g1SOd} Calculation of $\g_1$}
\end{boldmath}
The one-loop calculation is very similar to the $\SU(4)$ case.
The $\SO(d)$ Feynman rules are given in App.~\ref{SOdFrules}.
Again the $Z$-factor of the hyperbaryon takes the form of Eq.~(\ref{Z}).
This means  that the anomalous dimension depends
on which 4-dimensional Dirac tensor structure is used
(see Table~\ref{tabdirac}), but not on further details of the construction.
The $\SO(d)$ group theoretic factors are of course different.
For some details on their calculation, see App.~\ref{SOdidns}.
The results for the one-particle irreducible graphs are given in
the rightmost column of Table~\ref{tabclr},
and for the wave functions of the external legs in Table~\ref{tabCTO}.

The usual conventions for $SU(N)$ and $\SO(d)$ (see Appendices \ref{SUN}
and~\ref{apSOd}) are such that,
for $\SO(6)$, all group theoretic factors are bigger by a factor 2
than those of $\SU(4)$.  To account for this,
the coupling constants are to be related according to
$g[\SU(4)]=\sqrt{2}g[\SO(6)]$.

Using Eq.~(\ref{Z}) and Tables~\ref{tabCTO}, \ref{tabdirac} and~\ref{tabclr}
we finally obtain the following one-loop anomalous dimension
\begin{equation}
  \g_1 = \left\{ \begin{array}{ll}
    3(d-1)(d/4-1)\ , \qquad & P_1 \otimes P_2 \ , \\
    3(d-1)\ , & P_1 \g_\m \otimes P_2 \g_\m    \ , \\
    (d-1)(5-d/4)\ , & \s_{\m\n}\otimes\s_{\m\n} \ .
  \end{array} \right.
\label{g1all}
\end{equation}
This result is valid for all the $\SO(d)$ hyperbaryons we have discussed.

\section{\label{disc} Discussion}
The results we have presented are entirely straightforward.
Yet, to our knowledge, this is the first calculation of
the scaling dimension of any partial compositeness operator,
in a four dimensional, field theoretic completion of a Composite-Higgs
scenario.

By far, the literature on Composite Higgs  makes use of effective field
theories. (Compare the discussion in \cite{Giudice:2007fh,Contino:2013kra,Buchalla:2014eca} and in the review articles
\cite{Perelstein:2005ka,Contino:2010rs,Bellazzini:2014yua,Panico:2015jxa}.)
  These can be higher-dimensional theories, which are not
renormalizable; or they can be four-dimensional non-linear
sigma models, generalizing the chiral lagrangian of QCD.
The vast scope of this body of work shows that much can be learned
from the effective field theory approach.  Still, there are many questions
that can only be answered within an ultraviolet completion.
The values of anomalous dimensions are one example.
Following Ref.~\cite{FerKar,ferretti},
we have taken the notion of an ultraviolet completion to mean
an asymptotically-free gauge theory with fermionic matter.
These hypercolor models contain a candidate composite Higgs field
arising as a (pseudo) NGB from chiral symmetry breaking.
In addition, they contain hyperbaryons that can serve as top partners,
thereby allowing for the scenario of a partially composite top quark.

A key question for the viability of partial compositeness is whether
the ultraviolet scale $\L_{EHC}$ introduced in Eq.~(\ref{EHC}) can be taken
much higher than the hypercolor scale $\L_{HC}$ so that flavor constraints
are satisfied, while, at the same time, the top-quark coupling
to the hyperbaryon is sufficiently enhanced relative to its naive magnitude,
so that a realistic top-quark mass emerges.  Apart from noting that
a large anomalous dimension is not in conflict with rigorous bounds,
the effective field theory approach has nothing concrete to say
about this question.

In order to establish scaling properties, a study of the dynamics of
a concrete model is required.  As a first step, we have constructed
in this paper the candidate top-partner hyperbaryon operators in various
hypercolor models, and calculated their one-loop anomalous dimension.
Our main results are Eqs.~(\ref{g1PP}),~(\ref{g1gg}) and~(\ref{g1all}).
The encouraging aspect of these results is that in essentially all cases
that we studied, we find that quantum effects enhance,
rather than suppress, the coupling between the top quark and the hyperbaryon.

However, quite clearly, many of  the models of Ref.~\cite{FerKar}, with their minimal
fermion content, would not give rise to nearly as much enhancement
as one needs. (This includes the $\SU(4)$ (equivalently $\SO(6)$) model,
and the $\SO(8)$ model, the ones which maintain asymptotic freedom for QCD.)
The reason is that it is almost certain that the running
gauge coupling of these theories is QCD-like.  In QCD, we enter
the perturbative regime at about 0.1 Fermi.  The running on any shorter
distance scale is then dominated by one loop.  Experience with QCD is
that any further non-perturbative running from 0.1 Fermi to 1 Fermi
would produce at most some $O(1)$ additional enhancement.
In the hypercolor theory, the combined effect of one-loop running
from $\L_{EHC}$ to, say, $10\L_{HC}$, for which
the enhancement factor~(\ref{logenhnc}) is logarithmic,
plus some $O(1)$ additional enhancement
for the range of $10\L_{HC}$ to $\L_{HC}$, would not come anywhere close
to the total enhancement needed to overcome the power suppression
arising from naive dimensional counting.

Additional fermions could be added to the minimal fermion content
of the models of Ref.~\cite{FerKar}
for the purpose of slowing down the running.  As long as the coupling
is sufficiently weak, the relevant one-loop formula, Eq.~(\ref{enhnc}), is expected
to provide a reasonable approximation.  Of course, when the coupling
gets bigger, eventually perturbative calculations cannot be trusted
any more, and only a full-fledged non-perturbative lattice calculation
can determine the anomalous dimension.

Finally, we would like to comment that if it is assumed that
the scale $\L_{EHC}$ is only involved in the mass generation
for the top (or top and bottom) quarks,
while some other physics is responsible for mass generation for
the four lighter quarks and for the leptons,
then the flavor constraints are significantly weaker.  In this case
the ratio $\L_{EHC}/\L_{HC}$ can be taken much smaller than is usually assumed.
An important phenomenological constraint, which appears to be satisfied
quite comfortably by the hypercolor models,
comes from the $Z\to b\bar{b}$ decay \cite{ferretti,ACDP}.
To our knowledge, a more systematic study of the flavor constraints
of this scenario is not available to date.

\vspace{3ex}
\noindent {\bf Acknowledgments}
\vspace{3ex}

We would like to thank Ethan Neil for correspondence at the beginning
of this project.
This work was supported in part by the U.~S. Department of Energy,
and by the Israel Science Foundation under Grant no.~449/13.

\appendix
\begin{boldmath}
\section{\label{SUN} $\SU(N)$ groups}
\end{boldmath}
The group generators are hermitian, and satisfy the commutation relations
$[T_a,T_b] = i f_{abc} T_c$, where the structure constants $f_{abc}$
are fully antisymmetric.  Given an \irrep\, $r$, the
quadratic Casimir $C_2(r)$ is defined by $T_a T_a = C_2 I$ where $I$ is the
identity matrix, and the trace $T(r)$ by $\tr(T_a T_b) = T(r)\, \d_{ab}$.
These invariants are related by
\begin{equation}
  \frac{C_2(r)}{T(r)} = \frac{D({\rm adj})}{D(r)}  \ ,
\label{C2T}
\end{equation}
where $D(r)$ is the dimension of the representation.  See Table~\ref{tabCT}
for their values for some \irreps.

\begin{table}[t]
\vspace*{2ex}
\begin{center}
\begin{tabular}{ r | c c c } \hline \hline
     & $D$ & $T$ & $C_2$ \\ \hline 
fund & $N$ & $\frac{1}{2}$ &  $\frac{N^2-1}{2N}$ \\ 
adj & $N^2-1$ & $N$ & $N$ \\
sym2 & $\frac{N(N+1)}{2}$ & $\frac{N+2}{2}$ & $\frac{(N+2)(N-1)}{N}$ \\
as2 & $\frac{N(N-1)}{2}$ & $\frac{N-2}{2}$ & $\frac{(N-2)(N+1)}{N}$  \\
\hline \hline
\end{tabular}
\end{center}
\begin{quotation}
\floatcaption{tabCT}{Dimensionality $D(r)$, quadratic Casimir $C_2(r)$,
and trace $T(r)$ for some $\SU(N)$ representations:
fundamental (fund), adjoint (adj), two-index symmetric (sym2),
and two-index antisymmetric (as2).}
\end{quotation}
\vspace*{-4ex}
\end{table}

For the one-loop calculation we need the explicit form of the group generators.
For the sextet (as2) \irrep, the infinitesimal transformation is given by
$\d_a\c_{AB}=i(T_a)_{AB}^{CD}\, \c_{CD}$, with
\begin{equation}
  (T_a)^{CD}_{AB} =
 \half \left( \d^C_A (T_a)^D_B - \d^C_B (T_a)^D_A
 - \d^D_A (T_a)^C_B + \d^D_B (T_a)^C_A  \right) \ .
\label{TAS2}
\end{equation}
where $(T_a)_A^B$ are the usual generators in the fundamental \irrep.
We also need the closure relation satisfied by these generators,
\begin{equation}
  (T_a)^B_A (T_a)^D_C
  = \half \left( \d^B_C \d^D_A - \frac{1}{N} \d^B_A \d^D_C \right) \ .
\label{clsrrel}
\end{equation}

For completeness, we also record the general form of the one- and two-loop
coefficients of the beta function (for any gauge group),
\begin{eqnarray}
  b_1 &=& \frac{11}{3}\, C_2({\rm adj}) - \frac{4}{3}\,\sum_r N_f(r)\, T(r) \ ,
\label{b12}\\
  b_2 &=& \frac{34}{3}\, C_2^2({\rm adj})
  -\sum_r N_f(r)\, T(r) \left(\frac{20}{3}\, C_2({\rm adj}) + 4C_2(r) \right)\ ,
\nonumber
\end{eqnarray}
where the sum is over the fermion representations,
and $N_f(r)$ is the number of Dirac fermions in the \irrep\ $r$
(a Majorana or Weyl fermion counts as one half Dirac fermion).

\bigskip
\begin{boldmath}
\section{\label{apSOd} $\SO(d)$ groups}
\end{boldmath}
We label the generators as $M_{ij}$
where $i,j=1,2,\ldots,d,$ and $M_{ji}=-M_{ij}$.  They are conventionally
antihermitian, and satisfy the commutation relations
\begin{equation}
  [M_{ij},M_{k\ell}]
  = \d_{i\ell}M_{jk} -\d_{ik}M_{j\ell} -\d_{j\ell}M_{ik} +\d_{jk}M_{i\ell} \ .
\label{SOdcomm}
\end{equation}
In the vector \irrep, the $\SO(d)$ generators are
\begin{equation}
  (M_{ij}^{\rm vec})_{k\ell} = \d_{ik} \d_{j\ell} - \d_{jk} \d_{i\ell}  \ .
\label{vecrep}
\end{equation}
The definitions of the invariants $C_2(r)$ and $T(r)$ given in App.~\ref{SUN}
may be applied to $iM_{ij}$, the hermitian version of the generators.
See Table~\ref{tabCTO} for the invariants of some \irreps.

\begin{table}[t]
\vspace*{2ex}
\begin{center}
\begin{tabular}{ r | c c c } \hline \hline
     & $D$ & $T$ & $C_2$ \\ \hline
vec & $d$ & $2$ &  $d-1$ \\
adj & $\frac{d(d-1)}{2}$ & $2(d-2)$ & $2(d-2)$ \\
chir & $2^{d/2-1}$ & $2^{d/2-3}$ & $\frac{d(d-1)}{8}$ \\
\hline \hline
\end{tabular}
\end{center}
\begin{quotation}
\floatcaption{tabCTO}{Dimensionality $D(r)$, quadratic Casimir $C_2(r)$,
and trace $T(r)$ for some $\SO(d)$ representations:
vector (vec), adjoint (adj), and chiral spinor (chir).
The latter applies only when $d$ is even.}
\end{quotation}
\vspace*{-4ex}
\end{table}

The groups $\SU(4)$ and $\SO(6)$ are isomorphic.  However, the conventional
normalizations of $\SU(N)$ and $\SO(d)$ generators are such that every
$\SO(6)$ group invariant is twice as big compared to the
corresponding $\SU(4)$ invariant.  As explained in Sec.~\ref{g1SOd},
we make up for this by postulating that $g[\SU(4)]=\sqrt{2}g[\SO(6)]$.

\bigskip
\begin{boldmath}
\subsection{\label{dDirac} Dirac algebra in $d=2n$ dimensions}
\end{boldmath}
The (euclidean) Dirac algebra
\begin{equation}
  \{\G_i,\G_j\} = 2 \d_{ij} \ , \qquad i,j=1,2,\ldots,d \ ,
\label{diracalg}
\end{equation}
is realized on $2^n\times 2^n$ matrices, where $n=d/2$.
The generalization of $\g_5$ is defined as
\begin{equation}
  \G_h = \eta_h \G_1 \G_2 \cdots \G_{2d}
\label{g5}
\end{equation}
(the subscript $h$ stands for ``handedness'').
The phase factor $\eta_h$ is chosen such that $\G_h$ is hermitian
and $\G_h^2=1$.

\medskip\noindent {\it Explicit construction}.
In order to verify various
properties we will need, it is useful to have an explicit construction
of the Dirac matrices as tensor products of the three Pauli matrices $\s_a$
and the $2\times2$ identity matrix $I$.  The iterative construction
works slightly differently in $d=4n$ and $d=4n+2$ dimensions.
Starting with the $d=4n$ sequence we have, for $d=4$,
\begin{eqnarray}
\label{gmd4}
  \G_1 & = & \s_1 \otimes \s_2 \,, \\
  \G_2 & = & \s_2 \otimes \s_2 \,, \NON
  \G_3 & = & \s_3 \otimes \s_2 \,, \NON
  \G_4 & = & \idn \otimes \s_1 \,, \NON
  \G_h \ = \
  \G_5 & = & \idn \otimes \s_3 \,.
\nonumber
\end{eqnarray}
For $d=8$,
\begin{eqnarray}
\label{gmd8}
  \G_1 & = & \s_1 \otimes \s_1 \otimes \s_1 \otimes \s_2  \,, \\
  \G_2 & = & \s_2 \otimes \s_1 \otimes \s_1 \otimes \s_2  \,, \NON
  \G_3 & = & \s_3 \otimes \s_1 \otimes \s_1 \otimes \s_2  \,, \NON
  \G_4 & = & \idn \otimes \s_2 \otimes \s_1 \otimes \s_2  \,, \NON
  \G_5 & = & \idn \otimes \s_3 \otimes \s_1 \otimes \s_2  \,, \NON
  \G_6 & = & \idn \otimes \idn \otimes \s_2 \otimes \s_2  \,, \NON
  \G_7 & = & \idn \otimes \idn \otimes \s_3 \otimes \s_2  \,, \NON
  \G_8 & = & \idn \otimes \idn \otimes \idn \otimes \s_1  \,, \NON
  \G_h \ = \
  \G_9 & = & \idn \otimes \idn \otimes \idn \otimes \s_3  \,.
\nonumber
\end{eqnarray}
For the $d=4n+2$ sequence we start with $d=2$,
\begin{eqnarray}
\label{gmd2}
  \G_1 & = & \s_1 \,, \\
  \G_2 & = & \s_2 \,, \NON
  \G_h \ = \
  \G_3 & = & \s_3 \,.
\nonumber
\end{eqnarray}
Next, $d=6$,
\begin{eqnarray}
\label{gmd6}
  \G_1 & = & \s_1 \otimes \s_1 \otimes \s_1  \,, \\
  \G_2 & = & \s_2 \otimes \s_1 \otimes \s_1  \,, \NON
  \G_3 & = & \s_3 \otimes \s_1 \otimes \s_1  \,, \NON
  \G_4 & = & \idn \otimes \s_2 \otimes \s_1  \,, \NON
  \G_5 & = & \idn \otimes \s_3 \otimes \s_1  \,, \NON
  \G_6 & = & \idn \otimes \idn \otimes \s_2  \,, \NON
  \G_h \ = \
  \G_7 & = & \idn \otimes \idn \otimes \s_3  \,,
\nonumber
\end{eqnarray}
and $d=10$,
\begin{eqnarray}
\label{gmd10}
  \G_1 & = & \s_1 \otimes \s_1 \otimes \s_1 \otimes \s_1 \otimes \s_1 \,, \\
  \G_2 & = & \s_2 \otimes \s_1 \otimes \s_1 \otimes \s_1 \otimes \s_1 \,, \NON
  \G_3 & = & \s_3 \otimes \s_1 \otimes \s_1 \otimes \s_1 \otimes \s_1 \,, \NON
  \G_4 & = & \idn \otimes \s_2 \otimes \s_1 \otimes \s_1 \otimes \s_1 \,, \NON
  \G_5 & = & \idn \otimes \s_3 \otimes \s_1 \otimes \s_1 \otimes \s_1 \,, \NON
  \G_6 & = & \idn \otimes \idn \otimes \s_2 \otimes \s_1 \otimes \s_1 \,, \NON
  \G_7 & = & \idn \otimes \idn \otimes \s_3 \otimes \s_1 \otimes \s_1 \,, \NON
  \G_8 & = & \idn \otimes \idn \otimes \idn \otimes \s_2 \otimes \s_1 \,, \NON
  \G_9 & = & \idn \otimes \idn \otimes \idn \otimes \s_3 \otimes \s_1 \,, \NON
\G_{10}& = & \idn \otimes \idn \otimes \idn \otimes \idn \otimes \s_2 \,, \NON
\G_h \ = \
\G_{11}& = & \idn \otimes \idn \otimes \idn \otimes \idn \otimes \s_3 \,.
\nonumber
\end{eqnarray}
From these examples it should be clear how to proceed for each sequence.

\medskip\noindent {\it Spinor representations}.
The antisymmetrized product of two Dirac matrices
\begin{equation}
  \S_{ij} = \frac{1}{4}\, [\G_i,\G_j] \ .
\label{spingen}
\end{equation}
satisfies the $\SO(d)$ commutation relations~(\ref{SOdcomm}).
We may use these matrices to reexpress the Dirac algebra~(\ref{diracalg}) as
\begin{equation}
  \G_i \G_j = \d_{ij} + 2 \S_{ij} \ ,
\label{diracSigma}
\end{equation}

The $2^n$ dimensional space
on which they act spans a Dirac-like reducible representation.
Introducing $d$-dimensional chiral projectors:
\begin{equation}
  \cp_\pm = \half (1\pm \G_h) \ ,
\label{Ppm}
\end{equation}
and noting that $[\S_{ij},\G_h]=0$, we may split
$\S_{ij} = \S_{ij}^+ + \S_{ij}^-$, where
$\S_{ij}^\pm = \cp_\pm \S_{ij} = \S_{ij} \cp_\pm$.
The generators $\S_{ij}^+$ and $\S_{ij}^-$ act on the irreducible
chiral representations of $\SO(2n)$.

We introduce block notation, where every $2^n\times 2^n$
Dirac-space matrix $X$ is written in $2\times2$ block form
corresponding to the projectors $\cp_\pm$,
\begin{equation}
  X = \left( \begin{array}{cc}
         X_{++} & X_{+-} \\
         X_{-+} & X_{--}
      \end{array} \right) \ .
\label{block}
\end{equation}
where each block is a $2^{n-1}\times 2^{n-1}$ matrix.

\begin{table}[t]
\vspace*{2ex}
\begin{center}
\begin{tabular}{ r | c|c|c|c } \hline \hline
 d & $C^T = \eta_1 C$ & $C\G_h=\eta_2\G_h C$
 & $(C\G_i)^T=\eta_3 C\G_i$
 & $(C\G_i\G_h)^T=\eta_4 C\G_i\G_h$ \\ \hline
 2 & $-$ & $-$ & $+$ & $+$ \\
 4 & $-$ & $+$ & $+$ & $-$ \\
 6 & $+$ & $-$ & $-$ & $-$ \\
 8 & $+$ & $+$ & $-$ & $+$ \\
\hline\hline
\end{tabular}
\end{center}
\begin{quotation}
\floatcaption{tabSigns}{Sign factors $\eta_1,\ldots,\eta_4$ occurring
in $d=2n$ dimensions in the relations defined at the top of
each column.  Notice that $\eta_3=-\eta_1$ and that $\eta_2\eta_3\eta_4=-1$.
In calculating $\eta_4$ we use $\G_h^T=\G_h$.
The periodicity is 8, so that the signs for $2n>8$ may be found
on the line that corresponds to $2n$ mod 8.}
\end{quotation}
\vspace*{-4ex}
\end{table}

\medskip\noindent {\it Charge conjugation}.
In any even dimension $d=2n$ one can find a charge conjugation matrix $\cc$
that satisfies
\begin{equation}
  \cc \G_i = - \G_i^T \cc \ , \hspace{10ex} i=1,2,\ldots,2n \ ,
\label{C}
\end{equation}
and $\cc^{-1} = \cc^\dagger = \cc^T$.  For the Dirac matrices we have constructed
explicitly, the charge conjugation matrices follow a simple pattern
($\e=i\s_2$)
\begin{subequations}
\label{ccong}
\begin{eqnarray}
  \cc(2) &=& \e \ ,
\label{ccong2}\\
  \cc(4) &=& \e \otimes \s_3 \ ,
\label{ccong4}\\
  \cc(6) &=& \e \otimes \s_3 \otimes \e \ ,
\label{ccong6}\\
  \cc(8) &=& \e \otimes \s_3 \otimes \e \otimes \s_3 \ ,
\label{ccong8}\\
  \cc(10) &=& \e \otimes \s_3 \otimes \e \otimes \s_3 \otimes \e \ ,
\label{ccong10}
\end{eqnarray}
\end{subequations}
and so on.  $\cc$ is antisymmetric for $d=2,4$ mod 8,
and symmetric for $d=0,6$ mod 8.
Also, $[\cc,\G_h]=0$ for $d=4n$, whereas $\{\cc,\G_h\}=0$ for $d=4n+2$.
We have tabulated the signs that occur in some basic relations
in Table~\ref{tabSigns}.

It follows from Eq.~(\ref{C}) that
\begin{equation}
  \cc \S_{ij} = - \S_{ij}^T \, \cc \ .
\label{CSig}
\end{equation}
In $d=4n+2$ dimensions we have $\{\cc,\G_h\}=0$, which implies that the two
chiral \irreps\ projected by $\cp_\pm$ are complex conjugates.
For $d=4n$, $\cc$ commutes with $\G_h$.  Taking into account the symmetry
properties of $\cc$ itself (see the first column of Table~\ref{tabSigns})
the chiral \irreps\ are real for $d=8n$ and pseudoreal for $d=8n+4$.

\bigskip
\begin{boldmath}
\subsection{\label{dDiracodd} Dirac algebra in $d=2n+1$ dimensions}
\end{boldmath}
When $d$ is odd we take the last Dirac matrix to be $\G_{2n+1}=\G_h$.
We have to distinguish between two families.
For $d=4n+3$, $\cc$ anticommutes with $\G_h$, and
Eq.~(\ref{C}) applies for $i=1,2,\ldots,2n+1$.
As a result, the construction of an $\SO(d)$ vector from a spinor bilinear
works as for even $d$ (except of course that the spinor \irrep\ is not
chiral any more).  For $d=4n+1$, Eq.~(\ref{C}) does not generalize
to the last Dirac matrix.  Instead, we may use
\begin{equation}
  (\cc\G_h) \G_i = + \G_i^T (\cc\G_h) \ , \hspace{10ex} i=1,2,\ldots,2n+1 \ .
\label{Cgh}
\end{equation}

\subsection{\label{SOdidns} Some identities}
Here we collect some useful identities.  First,
\begin{equation}
  \G_i \G_j \G_i = (2-d) \G_j \ .
\label{GgG}
\end{equation}
Next, using the above identity and Eq.~(\ref{diracSigma}) we have
\begin{equation}
  -\S_{ij} \G_k \S_{ij} = \frac{(d-2)^2-d}{4}\;\G_k \ ,
\label{SgS}
\end{equation}
which is used for the color factor $\cg(a)$ in Table~\ref{tabclr}.
Last, for the color factor $\cg(b,c)$, we use
an identity involving the generators of both the vector and spinor
\irreps,
\begin{equation}
  \G_\ell \S_{ij} (M_{ij}^{\rm vec})_{\ell k}
  = -\S_{ij} \G_\ell (M_{ij}^{\rm vec})_{\ell k}
  = (d-1) \G_k \ .
\label{SMid}
\end{equation}
where the vector \irrep's generators $(M_{ij}^{\rm vec})_{k\ell }$
are given by Eq.~(\ref{vecrep}).  In order to arrive at the group theoretic factors
in the rightmost column of Table~\ref{tabclr} we have to divide these results
by 2, because the unconstrained summation on the left-hand side amounts
to summing over each $\SO(d)$ generator twice.

\begin{boldmath}
\subsection{\label{SOdFrules} $\SO(d)$ Feynman rules}
\end{boldmath}
In order to avoid double counting,
it is convenient to label the \SO($d$) generators
by an ordered pair $[ij]$ where $1\le i < j\le d$.
The tree-level gluon propagator is then
\begin{equation}
  \svev{A_{\m[ij]}A_{\n[kl]}} = \frac{\d_{\m\n}\d_{ik}\d_{jl}}{p^2} \ ,
\label{gluonprop}
\end{equation}
where we have used the Feynman gauge for definiteness.
The (massless) fermion lagrangian is
\begin{eqnarray}
  \cl_F^\pm &=& \bj_{A\a}\, (\g_\m)_{\a\b} (D_\m)_{AB}^\pm\, \j_{B\b} \ ,
\label{LF}\\
  (D_\m)_{AB}^\pm &=& \d_{AB} \partial_\m
  + \sum_{i<j} A_{\m[ij]} (\S_{[ij]}^\pm)_{AB} \ ,
\label{corderF}
\end{eqnarray}
where the $\pm$ superscript refers to the chirality of the \irrep.
Notice the absence of a factor of $i$ in front of the gauge field's term
in Eq.~(\ref{corderF}), because the generators are already antihermitian.

When we contract a gluon propagator between two fermion vertices
with generators $M_{[ij]}^{(1)}$ and $M_{[ij]}^{(2)}$
we will get, schematically,
\begin{equation}
  \sum_{i<j} M_{[ij]}^{(1)} \otimes M_{[ij]}^{(2)}
  = \half \sum_{ij} M_{ij}^{(1)} \otimes M_{ij}^{(2)} \ ,
\label{twicesumb}
\end{equation}
where, in the unrestricted summation on the right-hand side,
we have used that $M_{ij}=-M_{ji}$ by definition.

\section{\label{dirac4} 4-dimensional Dirac matrices}
Denoting by $C$ the charge conjugation matrix in four dimensions,
Eq.~(\ref{ccong4}), the behavior under charge conjugation is
\begin{eqnarray}
\label{CCrules}
  (C\, I)^T &=& -(C\, I) \\
  (C\, \g_5)^T &=& -(C\, \g_5) \NON
  (C\, \g_\m)^T &=& +(C\, \g_\m) \NON
  (C\, \g_5\g_\m)^T &=& -(C\, \g_5\g_\m) \NON
  (C\, \s_{\m\n})^T &=& +(C\, \s_{\m\n}) \ ,
\nonumber
\end{eqnarray}
where in this appendix, $I$ is the identity matrix in Dirac space.
As usual (compare Eq.~(\ref{spingen}))
\begin{equation}
  \s_{\m\n} = (i/2) [\g_\m,\g_\n] \ .
\label{sSmn}
\end{equation}
When working out the one-loop diagrams, thanks to the common color factor
of the graphs in Figs.~\ref{diags}(b) and~(c),
summing them together always simplifies the Dirac algebra.
For the derivation of the results in the second row of Table~\ref{tabdirac}
we use the following identities.  First,
\begin{equation}
  \g_\m \g_\n \otimes (\g_\m \g_\n + \g_\n \g_\m) = 8 I \otimes I \ ,
\label{2and2}
\end{equation}
which follows immediately from the Dirac algebra.  The next identity
\begin{equation}
  \g_\m \g_\n \g_\r \otimes (\g_\m \g_\n \g_\r + \g_\r \g_\n \g_\m)
  = 20 \g_\r \otimes \g_\r \ ,
\label{3and3}
\end{equation}
can be proved using
\begin{equation}
   \g_\m \g_\n \g_\r = \d_{\m\n} \g_\r - \d_{\m\r} \g_\n + \d_{\n\r} \g_\m
   + \e_{\m\n\r\t} \g_5 \g_\t \ .
\label{ggg4}
\end{equation}
Finally,
\begin{eqnarray}
  && \hspace{-10ex}
  \g_\m \g_\n \s_{\l\r} \otimes (\g_\m \g_\n \s_{\l\r} + \s_{\l\r} \g_\n \g_\m)  \ =
\label{sgg}\\
  &=& (\d_{\m\n}-i\s_{\m\n}) \s_{\l\r} \otimes
  [(\d_{\m\n}-i\s_{\m\n}) \s_{\l\r} + \s_{\l\r} (\d_{\m\n}+i\s_{\m\n}) ]
\NON
  &=& 8 \s_{\l\r} \otimes \s_{\l\r}
      - \half [\s_{\m\n},\s_{\l\r}] \otimes [\s_{\m\n},\s_{\l\r}]
\NON
  &=& 24 \s_{\l\r} \otimes \s_{\l\r} \ ,
\nonumber
\end{eqnarray}
where we have used $\s_{\m\n} = 2i \S_{\m\n}$
and Eqs.~(\ref{SOdcomm}) and~(\ref{diracSigma}).

\vspace{5ex}

\end{document}